# Visualizing Temperature-Dependent Phase Stability in High Entropy Alloys

Daniel Evans[1], Jiadong Chen[1], Geoffroy Hautier[2], Wenhao Sun[1*]

[1]Department of Materials Science and Engineering, University of Michigan, Ann Arbor, MI, 48109, United States
[2]Thayer School of Engineering, Dartmouth College, Hanover, NH, 03755, United States

*Correspondence to whsun@umich.edu

## Abstract

High Entropy Alloys (HEAs) contain near equimolar amounts of five or more elements and are a compelling space for materials design. Great emphasis is placed on identifying HEAs that form a homogeneous solid-solution, but the design of such HEAs is hindered by the difficulty of navigating stability relationships in high-component spaces. Traditional phase diagrams use barycentric coordinates to represent composition axes, which require $D = (N – 1)$ spatial dimensions to represent an $N$-component system, meaning that HEA systems with $N > 4$ components cannot be readily visualized. Here, we propose forgoing barycentric composition axes in favor of two energy axes: a formation-energy axis and a 'reaction energy' axis. These *Inverse Hull Webs* offer an information-dense 2D representation that successfully capture complex phase stability relationships in $N \geq 5$ component systems. We use our new diagrams to visualize the transition of HEA solid-solutions from high-temperature stability to metastability upon quenching, and identify important thermodynamic features that are correlated with the persistence or decomposition of metastable HEAs.

**Introduction**

Traditional alloy design consists of one or two principal elements alloyed with small amounts of supplementary elements. On the other hand, High Entropy Alloys (HEAs) contain near equimolar amounts of five or more elements and offer promising mechanical behavior [1, 2, 3, 4, 5], corrosion resistance [6, 7], and compositional tunability [8, 9]. Additionally, high-component ceramics [10, 11, 12, 13] can be semiconductors [14, 15] or used as cathodes in Li-ion batteries [16, 17], illustrating the promise of entropy-stabilized materials. The large number of compositional degrees of freedom available in the search for high-component materials gives rise to an exciting and still largely-unexplored materials design space [18]. So far, significant emphasis has been placed on identifying the single-phase and multi-phase equilibrium regions in an HEA phase diagram [19, 20, 21]—where a single-phase HEA is defined by random mixing of all atoms on the same lattice in a homogeneous solid-solution. Thermodynamically, this single-phase HEA solid-solution will be stable if it is lower in free energy than any linear combination of its lower-component ordered intermetallics and solid-solutions.

Unfortunately, there is a lack of suitable tools to visualize thermodynamic competition in high-component chemical space, which impedes the guided exploration of HEAs. Traditional phase diagrams use barycentric composition axes, where compound compositions are given by the lever rule. A barycentric representation of an $N$-component system requires $D = (N – 1)$ spatial dimensions, meaning systems with $N \geq 5$ components cannot be visualized in $D \leq 3$ spatial dimensions. Past attempts to visualize HEA stability include projecting a five-component phase diagram into three dimensions, or constructing a series of quaternary phase diagrams [22]. However, these approaches do not scale to an arbitrary number of components, and do not quantitatively display relative thermodynamic relationships between phases—especially as a function of temperature. Developing a scheme to visualize stability relationships in high-component systems would help facilitate the discovery of promising HEA compositions.

Here, we propose a new information-dense [23, 24] 2D representation that captures the essence of phase stability relationships in high-component systems, while remaining easily interpretable. We choose to forgo barycentric composition axes in order to avoid the scaling relationship between components and dimensions; instead adopting a graph-based representation of stability relationships. However, unlike previous graph-based stability representations which do not have meaningful $x$ and $y$-axes [25, 26], we assign a formation energy axis and a reaction energy axis, capturing the absolute and relative stabilities of competing $N$-component phases. Furthermore, we design a variety of plot features including color, line-width, and marker shape to retain salient compositional information lost by eliminating barycentric composition axes.

We name these diagrams *Inverse Hull Webs*, which we use to illustrate temperature-dependent phase-stability during the quenching of HfMoNbTiZr and AlCrFeNi, which are experimentally reported to be a single-phase and a multiphase HEA system, respectively. Our Inverse Hull Webs successfully capture high-level trends in HEA solid-solution stability, as well as precise thermodynamic details regarding phase decomposition across the HEA composition space. Additionally, by animating our diagrams as a function of temperature, we can track the transition of an HEA solid-solution from stability to metastability and identify which intermetallic compounds threaten the thermodynamic stability of the HEA solid-solution phase.

More broadly speaking, our use of energy axes allows our Inverse Hull Webs to offer a complementary perspective to traditional phase diagrams, highlighting important thermodynamic features that are typically missing from traditional phase diagrams such as convex hull depth, reaction driving forces, metastability, and tendency for phase separation. Our visualization is therefore useful for generally understanding materials stability, including in ordered, low-component materials systems. Beyond Inverse Hull Webs, we believe that there may exist many new creative ways to visualize phase stability relationships, which can serve as new theoretical tools to discover and design advanced novel materials.

**Results**

**Design of Inverse Hull Webs**

Phase stability is governed by the convex hull construction, as illustrated in **Figure 1a**. By plotting the formation energy of materials against composition, the stable phases lie on the lower convex hull formed between compounds and their elemental reference states [27]. Thermodynamically stable phases are vertices on the convex hull, while metastable phases have an energy above the convex hull at their associated composition [28]. Phase coexistence occurs when a point lies on a facet, or simplex, of the convex hull. The convex hull of an *N*-component system is plotted in a barycentric representation in (*N*-1) dimensions. The scaling relationship between the number of components, *N*, and the number of spatial dimensions, *D*, is the source of the difficulty in visualizing stability relationships in high-component systems. Notably, a 5-component convex hull would need to be visualized in 4 spatial dimensions.

Here, we propose a new 2D visualization scheme, which we name *Inverse Hull Webs*, that captures important information from the thermodynamic convex hull and does not use barycentric composition axes, meaning that our diagrams are not spatially constrained by the number of components in the system. **Figure 1b** illustrates the conversion of a traditional convex hull to our Inverse Hull Webs for the binary Al-Fe system at 1250K.

We use formation energy on the *y*-axis and *inverse hull energy* on the *x*-axis of our Inverse Hull Webs. Formation energy is the free energy difference between a phase and the linear combination of its corresponding pure elements, as noted in **Figure 1** by an arrow for AlFe$_3$. In the computational materials design community, it is conventional to define energies relative to the convex hull; a stable phase has $\Delta E_{hull}$ = 0 and metastable phases have an energy 'above the hull', which is the difference in formation energy between a metastable phase and its stable decomposition products [28]. Here, we generalize this concept—defining the *inverse hull energy* to be the reaction energy to a phase from its stable neighbors in composition space—which we call the *hull reactants*. In other words, the inverse hull energy for a metastable phase is simply its 'energy above the hull', but for a stable phase it is the energy *below* a hypothetical convex hull where the stable phase was removed, as illustrated for Al$_6$Fe in **Figure 1**. Stable and metastable phases can be quickly identified on an Inverse Hull Web, as they fall to the left or right of *x* = 0, respectively.

By eliminating barycentric axes, we lose important information regarding the composition of the phases. We design a variety of plot features—such as arrows, arrow width, marker shape, and color—to regain this compositional information. First, we use arrows to connect hull reactants to product phases, based on the inverse hull energy definition. We emphasize that these arrows do *not* represent tie lines, for example as was used by Hegde *et al.* [25] in their phase stability network representation. While tie lines visualize two-phase coexistence, our arrows indicate reactions between compositional neighbors on the convex hull. This distinction becomes clear in $N \geq 3$ component systems, as the 'hull reactants' are not necessarily the phases connected to a compound by its tie lines.

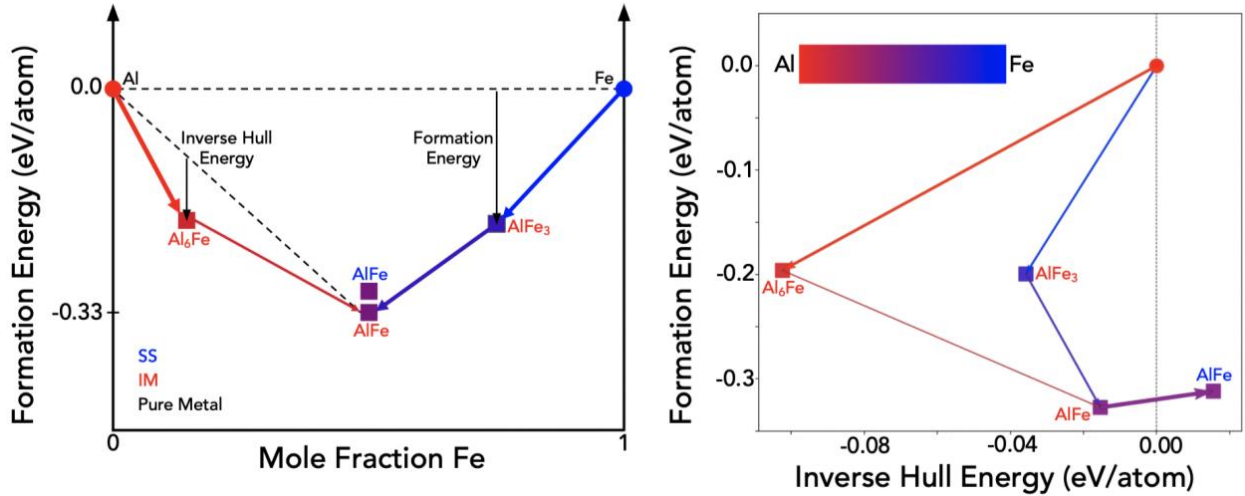

**Fig 1. Conversion of convex hull to Inverse Hull Web.** (**Left**) Convex Hull of Al-Fe system at 1250K, with formation energy on the *y*-axis and mole fraction Fe on the *x*-axis. The dotted line connecting Al and AlFe demonstrates the process for finding $Al_6Fe$'s hull reactants and inverse hull energy. The arrow width representation of phase fraction is shown as the arrow connecting $Al_6Fe$ to Al is thicker than the arrow connecting $Al_6Fe$ to AlFe. (**Right**) Inverse hull web of Al-Fe system at 1250K, with formation energy on the *y*-axis and inverse hull energy on the *x*-axis. The AlFe solid-solution lies to the right of the dotted line, indicating it is metastable with respect to an AlFe intermetallic.

We retain the coefficients of a stoichiometrically-balanced reaction by using arrow widths to represent the phase fraction of each hull reactant, where thicker arrows correspond to larger phase fractions. We use text color to represent phase type and marker shape to represent the number of elements in the compound. Circles represent pure elements, squares represent binary compounds, triangles represent ternary compounds, diamonds represent quaternary compounds, and pentagons represent quinary compounds. In our study of HEAs, blue text indicates a solid-solution, red text indicates an ordered intermetallic, and black text indicates a pure element.

We use color to represent the composition of each phase [29, 30], shown by the color bar in **Figure 1**, where Al is red and Fe is blue, and AlFe is purple. Our color scheme is based off of barycentric coordinates, where each vertex of an (*N* – 1) dimensional simplex in an *N*-component compound is assigned a color. Meyer *et al.* proposed a method to project an *N*-dimensional simplex into a regular *N*-gon, allowing us to construct a 2-dimensional barycentric color legend [31], even for high-component systems. This is accomplished using Equation 1 below to calculate a 2D position corresponding to a set of vertex weights.

$$w_j = \frac{\cot(\theta_{j+1}) + \cot(\theta_{j-1})}{|p - v_j|^2} \qquad (1)$$

where $w_j$ refers to the barycentric weight of the $j^{th}$ vertex, $v_j$ refers to the 2D position of the $j^{th}$ vertex, p refers to the 2D position of the point of interest, and $\theta_{j\pm1}$ refers to the angle between the lines connecting p to $v_j$ and p to $v_{j\pm1}$. **Figure 2** demonstrates this barycentric projection with the stable phases in the Al-Cu-Co-Ni-Fe system at 1500K, where each element lies on the vertex of a pentagon and all compounds lie in between.

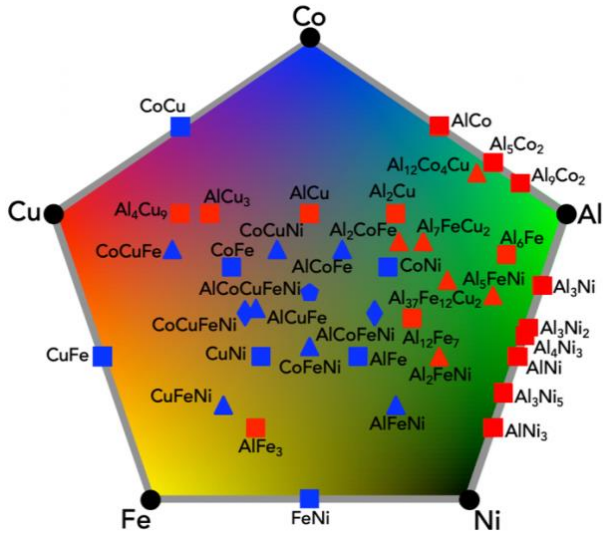

**Fig 2. Projected Barycentric Phase Diagram.** 4D phase diagram of the Al-Cu-Co-Ni-Fe system at 1500K projected into a pentagon, where each vertex represents an element. A 2D point's distance from each vertex is related to the composition corresponding to that point. Binary tie lines connect vertices, lying along edges of the polygon for adjacent elements and extending across the polygon for non-adjacent elements. Ternary compositions lie within a triangle connecting three elements, and quaternary compositions lie within a quadrilateral.

One drawback of the 2D *N*-gon barycentric projection is that a composition do not have a unique 2D location in the *N*-gon. For example, CoNi and $Al_2FeCo$ are at similar locations in the pentagon but represent very different compositions in **Figure 2**. This limitation arises from the projection of a 5-dimensional composition space into 2-dimensions and is unfortunately unavoidable. However, oftentimes the most stable ordered intermetallics in a system are clustered in a subregion of composition space, for example, as indicated by the high density of intermetallics (red squares) near Al in **Figure 2**. This means that high-level compositional threats to HEA solid-solution stability can often still readily be identified with this colorimetric representation.

**Application of Inverse Hull Webs to HEAs**

HEAs are entropically-stabilized at high temperature by the configurational entropy that arises from mixing many elements on a single disordered lattice. A major goal in HEA development is to identify compositions that can form single-phase solid-solutions. Thermodynamically, the configurational entropy and Gibbs free energy of solid-solution and intermetallic phases can be found in Equations 2 and 3, respectively, where $x_i$ refers to the molar fraction of the $i^{th}$ component.

$$\Delta S_{SS} = -k_B \sum_{i=1}^{N} x_i \ln(x_i) \, ; \, \Delta S_{IM} = 0 \tag{2}$$

$$\Delta G = \Delta H - T\Delta S \tag{3}$$

Here, we calculate the formation enthalpies of solid-solution phases using DFT-calculated regular solution parameters from Bokas *et al.* [32], and use the formation enthalpies of ordered intermetallics as calculated in the Materials Project [33]. We assume that the configurational entropy of all solid solutions is ideal and that differences in vibrational entropy per atom between solid phases is negligible at high temperatures, as can be expected from the law of Dulong-Petit [34].

The $\Delta S$ for a 5-component solid-solution is 1.6R, meaning the T$\Delta S$ contribution at 1000K is -0.14 eV/atom. However, formation energies of ordered binary intermetallics are often more negative than -0.3 eV/atom, meaning that below 1000K there is a high probability that intermetallics become thermodynamically favored over solid-solutions with high configurational entropy. This means that many single-phase HEAs may be metastable at room temperature relative to these ordered lower-component intermetallics. Thus, many single-phase HEAs should be temperature-stabilized somewhere between room temperature and the melting point of the alloy.

We will use our Inverse Hull Webs to visualize the transition of an HEA solid-solution from stability to metastability as it is quenched and illustrate how it can be used to guide the design of compelling HEA compositions. We will examine two HEAs: HfMoNbTiZr, which was experimentally-reported to be single-phase HEA when quenched; and AlCrFeNi, which phase-separates upon quenching. In our analysis, the only metastable phases we consider are solid-solution phases, and not metastable intermetallics. We display Inverse Hull Webs of these systems at a series of representative temperatures during quenching.

**Inverse hull web of an experimentally-observed single-phase HEA**

First, we examine HfMoNbTiZr, an HEA that Tseng *et al.* found be to be a single BCC solid-solution phase at room temperature after being cast through vacuum arc melting and remelted at least four times [36]. The Inverse Hull Web of HfMoNbTiZr is shown in **Figure 3** at four important temperatures during a quenching process; first, at the Mo melting temperature, 2896K; and lastly, at room temperature, 300K. We also show the Inverse Hull Web at two critical temperatures that we have identified during quenching.

We call the first critical temperature the 'critical solid-solution' temperature, or the lowest temperature where the HEA solid-solution is stable, which is 563K in this system. This temperature represents where the HEA solid-solution becomes metastable during quenching, giving insight into the likelihood the alloy will be found as a single phase at room temperature. The next critical temperature we consider is the alloy's 'critical adjacent phase' temperature, or the lowest temperature where the *N*-component HEA solid-solution phase's hull reactants are all of the compositionally-adjacent, (*N* – 1) component solid-solutions, which is 1167K in this system. This serves as a proxy for the temperature where enthalpic effects in the competing solid-solution and ordered phases become important.

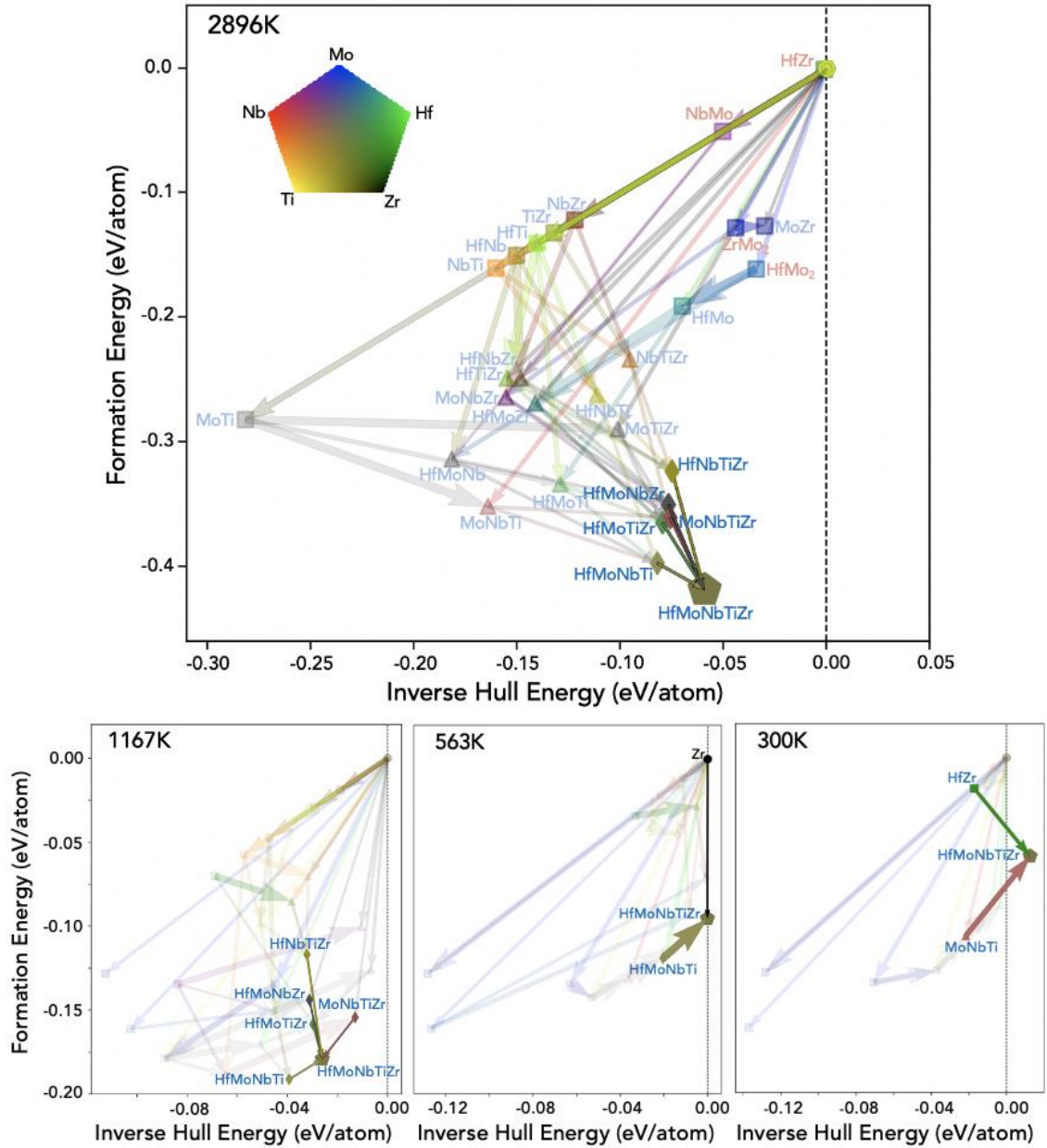

**Fig 3. Inverse hull web of HfMoNbTiZr.** HfMoNbTiZr system at temperatures of 2896K, 1167K, 563K, and 300K. The top Inverse Hull Web is at 2896K, the highest melting temperature in the Hf-Mo-Nb-Ti-Zr system, and the bottom left Inverse Hull Web is at 1167K, HfMoNbTiZr's 'critical adjacent phase' temperature. The bottom left Inverse Hull Web is at 563K, HfMoNbTiZr's 'critical solid-solution' temperature, and the bottom right Inverse Hull Web is at room temperature. We leave the single solid-solution and its hull reactants opaque, while we make all other phases partially transparent.

In **Figure 3**, we see that intermetallics do not offer significant thermodynamic competition in the HfMoNbTiZr system; the binary intermetallics (in red text) all have formation energies more positive than -0.05 eV/atom, and there are no higher-component intermetallics. Out of all of the solid-solutions, MoTi is the only phase likely to pose a threat to HEA stability in the Hf-Mo-Nb-Ti-Zr system. Solid-solutions are stable across much of this system, indicated by the large number of blue marker labels, and three out of the four stable intermetallics contain Mo while the fourth, HfZr, has negligibly small formation energy. HfMoNbTiZr has the lowest formation energy in the entire 5-component chemical space, and at the Mo melting temperature, has a sizable inverse hull energy of -0.05 eV/atom.

As the system is quenched, HfMoNbTiZr's 'critical adjacent phase' temperature is 1167K, meaning enthalpic effects do not become important until roughly 1700K below Mo's melting temperature. The HEA solid-solution is stable until only a few hundred Kelvin above room temperature as it has a 'critical solid-solution' temperature of 563K. Additionally, the HEA solid-solution is only slightly metastable at room temperature with an energy above the hull of 12 meV/atom, a small energy value relative to the scale of metastability in quinary compounds [28]. Altogether, these features suggest that HfMoNbTiZr is likely to persist as a metastable single solid-solution at room temperature.

In **Figure 3**, we see that the HEA solid-solution initially decomposes into Zr and HfMoNbTi when it becomes unstable, indicated by the arrows extending from the pentagon at 563K in **Figure 3**. This means the quaternary HfMoNbTi solid-solution is more stable than the quinary HfMoNbTiZr alloy, suggesting that the addition of Zr to HfMoNbTi may give rise to a miscibility gap in the alloy at lower temperatures. Additionally, the quaternary HfMoNbTi solid-solution has the lowest formation energy in the system at 1167K, further suggesting the addition of Zr to HfMoNbTi increases the energy of the solid-solution.

Our Inverse Hull Webs display the low critical temperatures, lack of energetically deep intermetallic phases, and low metastability of the HEA solid-solution at room temperature in the Hf-Mo-Nb-Ti-Zr system, allowing us to infer that this system is amenable to persist as a metastable single solid-solution HEA at ambient conditions. These stability relationships corroborate the experimental observations of Tseng *et al* of a room temperature solid-solution in HfMoNbTiZr, demonstrating the utility of our new diagrams.

**Inverse hull web of an experimentally-observed multiphase HEA**

Our Inverse Hull Webs can also be used to visualize features that may inhibit single solid-solution stability in HEAs. We demonstrate this with AlCrFeNi, a derivative of the CoCrFeMnNi Cantor alloy achieved by replacing Co and Mn with Al. The Cantor alloy was one of the first HEAs found to exhibit a single solid-solution microstructure, however it was later discovered that the HEA solid-solution phase becomes metastable at around 800°C [18, 19]. Chen *et al* found the AlCrFeNi alloy to be a phase separated AlNi B2 intermetallic and CrFe BCC solid-solution at room temperature after being cast through arc melting and remelted five times [37].

We show the AlCrFeNi Inverse Hull Web in **Figure 4** at 3853K, 2180K, 1650K, and 300K, corresponding to AlCrFeNi's 'critical adjacent phase' temperature, Cr's melting temperature, AlCrFeNi's 'critical solid-solution' temperature, and room temperature.

**Fig 4. Inverse hull web of AlCrFeNi.** AlCrFeNi system at temperatures of 2180K, 3853K, 1650K, and 300K. The top Inverse Hull Web is at 2180K, the highest melting temperature in the Al-Cr-Fe-Ni system, and the bottom left Inverse Hull Web is at 3853K, AlCrFeNi's 'critical adjacent phase' temperature. The bottom left Inverse Hull Web is at 1650K, AlCrFeNi's 'critical solid-solution' temperature, and the bottom right Inverse Hull Web is at room temperature.

**Figure 4** shows that Al-Ni intermetallics threaten HEA solid-solution stability in all alloy systems containing these elements. Al-Ni intermetallics have very negative formation energies of -0.6 to -0.7eV, as well as small inverse hull energies of -0.01 to -0.05eV at all casting-relevant temperatures, which is visually apparent by the large number of green-yellow markers in the lower right of every Inverse Hull Web in **Figure 4**. This suggests that the Al-Ni isopleth is littered with competing low-energy compounds.

Upon casting, the stability of the HEA solid-solution is immediately threatened by Al-Ni intermetallics. The highest melting temperature in the Al-Cr-Fe-Ni system is 2180K, while the 'critical adjacent phase' temperature is 3853K. In this alloy, the 'critical adjacent phase' temperature is the lowest temperature where the 4-component solid-solution's hull reactants are each of the 3-component solid-solutions, meaning below this temperature the single solid-solution begins to compete with either binary solid-solutions or ordered intermetallics. **Figure 4** shows that the HEA solid-solution has a similar formation energy as Al-Ni intermetallics at 3853K, suggesting these intermetallics begin to threaten the stability of the HEA solid-solution at a very high temperature.

The HEA solid-solution in this system then becomes metastable at 1650K, 600K higher than in the Cantor alloy and more than 1000K higher than in HfMoNbTiZr. Additionally, the HEA solid-solution is highly metastable at room temperature with an energy above the hull of 0.15 eV/atom, a large energy value relative to the scale of metastability in quaternary compounds and more than 0.10 eV/atom more metastable than HfMoNbTiZr at room temperature [28].

Altogether, AlCrFeNi's high 'critical adjacent phase' temperature relative to its constituents' melting temperatures, high 'critical solid-solution' temperature, and high metastability at room temperature are characteristic of phase-separating HEA systems. Even though this system is only a quaternary HEA, the threat posed by low-energy Al-Ni binary intermetallics should threaten other HEAs that contain these two elements. Our Inverse Hull Webs rationalize the experimental observations of Chen *et al* of phase separation in this system into an AlNi intermetallic and a CrFe solid-solution.

**Analysis of critical temperatures in HEAs**

The *critical solid solution* and *critical adjacent phase* temperatures of an HEA are features that may be more broadly indicative of HEA solid-solution stability. We next calculate these critical temperatures for 103 equimolar, as-cast HEAs found in Reference [35] and use each temperature individually to classify the HEAs as single or multiphase. **Figure 5** below shows the critical temperatures of each HEA divided by the highest melting temperature in the HEA system, along with the classification accuracy resulting from choosing a single value to distinguish between single and multiphase. Relative 'critical solid-solution' temperature is shown on the left and relative 'critical adjacent phase' temperature is shown on the right.

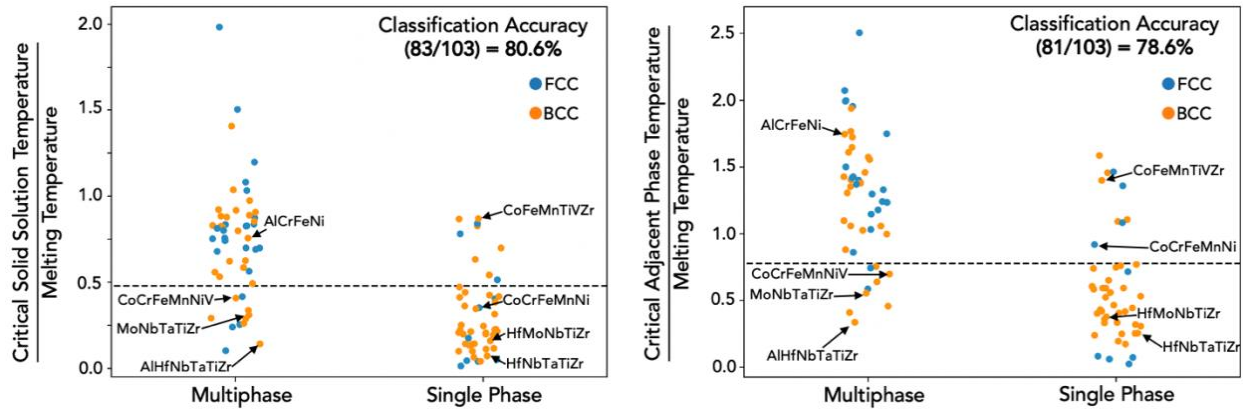

**Fig. 5 Critical Temperatures of 103 equimolar, as-cast HEAs.** (**Left**) 'Critical solid-solution' temperature of each HEA divided by the highest melting temperature in the HEA system. (**Right**) 'Critical adjacent phase' temperature of each HEA divided by the highest melting temperature in the HEA system. Only FCC and BCC phases are considered, as regular solution parameters in [32] are for FCC and BCC phases. The dotted line is the cutoff value yielding the highest classification accuracy when distinguishing between single and multiphase HEAs. Values for AlCrFeNi, CoCrFeMnNi, HfMoNbTiZr, HfNbTaTiZr, MoNbTaTiZr, AlHfNbTaTiZr, CoFeMnTiVZr, and CoCrFeMnNiV are labelled with arrows.

**Figure 5** above shows that both 'critical solid-solution' temperature and 'critical adjacent phase' temperature are effective at distinguishing between single and multiphase HEAs, achieving classification accuracies of 80.6% and 78.6%, respectively, even when used in isolation and selecting a single cutoff value. The dotted line in the "critical solid-solution" temperature plot is slightly below 0.5, suggesting that the single solid-solution needs to be stable at half of the casting temperature for the HEA to be a single solid-solution at room temperature. Additionally, the dotted line in the 'critical adjacent phase' temperature plot is roughly 0.8, suggesting entropic effects need to dominate down to 80% of the casting temperature for an HEA to be a single solid-solution at room temperature. AlCrFeNi and HfMoNbTiZr were correctly classified with both temperatures, and the Cantor alloy was as well. We identified these features here by physical intuition, but they may also serve as valuable new features for machine-learning classification models [38, 39, 40, 41] to predict which HEA compositions are likely to persist as single solid-solutions [42, 43, 44, 45].

**Discussion**

Our Inverse Hull Webs display both high-level trends in temperature-dependent phase stability as well as precise details regarding the thermodynamic competition between compounds. Importantly, by displaying salient thermodynamic information in 2D for any $N$-component system, we overcome a major limitation of traditional phase diagrams for analyzing high-component systems—as shown here to rationalize solid-solution stability in HEAs. Regions of composition space that pose a threat to HEA solid-solution stability can be identified by compounds with very negative formation energies, such as Al-Ni intermetallics in the AlCrFeNi system. The thermodynamic incentive for an HEA to persist as a solid-solution or to decompose can assessed by its temperature-dependent inverse hull energy. Altogether, these Inverse Hull Webs can serve as valuable new tools to assist in the guided discovery of experimentally-synthesizable disordered (or ordered) high-component materials.

Our analysis of single solid-solution stability in these high-entropy alloys further reveals two new important features—the 'critical solid-solution' temperature and 'critical adjacent phase' temperature. 'Critical solid-solution' temperature is the lowest temperature where the HEA solid-solution is stable and thus estimates where the HEA solid-solution begins to decompose. 'Critical adjacent phase' temperature is the lowest temperature where the $N$-component HEA solid-solution's hull reactants are each of the ($N$-1) component solid-solutions, estimating where entropic effects no longer dominate. These features may illuminate which HEA compositions are likely to be found as a single solid-solution phase at room temperature and may be valuable features for a machine-learning model aimed at predicting single phase HEA compositions.

Generally speaking, our Inverse Hull Webs offer a new thermodynamic perspective to complement the traditional representation of a convex hull. Typical phase diagrams visualize only the equilibrium phases at a given composition, and do not provide formation energy or reaction energy information. Our Inverse Hull Webs reveal other important aspects of the thermodynamic landscape, such as convex hull depth, reaction driving forces, metastability, and tendency for phase separation; all of which are difficult to ascertain from the traditional phase diagram. Even though our Inverse Hull Webs do not use barycentric composition axes, we are able to recover composition information by utilizing a variety of plot features, such as color, line width, and marker shape. We hope that by rethinking the classical approach to constructing phase diagrams, our work inspires the creation of new yet-to-be-designed visualizations that can illuminate important trends in the *synthesis-structure-property-performance* relationships of advanced materials.

## Methodology

### Calculation of compound free energies

Intermetallic free energies were obtained from The Materials Project database [33] through the *pymatgen* Python package [46]. Solid-solution free energies were calculated using a regular solution model as shown in Equation 4, where $N$ refers to the number of components, $x_i$ refers to the mole fraction of the $i^{th}$ element, and $\Omega_{ij}$ is the parameter that quantifies the interaction resulting from mixing the $i^{th}$ and $j^{th}$ elements.

$$\Delta G_{mix} = \sum_{i=1}^{N} \sum_{j>i}^{N} \Omega_{ij} x_i x_j + k_B T \sum_{i=1}^{N} x_i \ln(x_i) \qquad (4)$$

Bokas *et al* used DFT free energies of binary solid-solutions to fit a regular solution parameter $\Omega$ for each binary pair of elements out of 27 elements commonly used in HEAs [32]. These regular solution parameters were used to calculate solid-solution free energies in our Inverse Hull Webs. The lower free energy of the FCC and BCC solid-solutions was found for each equimolar combination of elements in the chemical system being plotted. All compound free energies were referenced to the stable phases of the pure elements composing the compound. Formation and inverse hull energies were calculated using *pymatgen* [46]. All Inverse Hull Webs were constructed using the matplotlib Python package [47].

Acknowledgments

The research was funded by the Walloon Region under the agreement no. 1610154- EntroTough in the context of the 2016 WalIinnov call. Computational resources were provided by the supercomputing facilities of the UCLouvain and the Consortium des Equipements de Calcul Intensif (CECI) en FederationWallonie Bruxelles, funded by the Fonds de la Recherche Scientifique de Belgique (FNRS).